\let\ifarxiv=\iftrue     % ARXIV VERSION
\pdfoutput=1

\ifarxiv

\documentclass[12pt,a4paper]{article}
\usepackage[a4paper,text={450pt,650pt},centering]{geometry}

\fi

\ifarxiv\else

\documentclass[11pt,a4paper]{article}
\usepackage{mathptmx}
\usepackage[a4paper,text={130mm,198mm}]{geometry}

\fi

\usepackage{amsmath,amssymb}
\usepackage[bookmarks=true,hyperfigures=true]{hyperref}
\usepackage{graphicx}
\usepackage[nosort]{cite}
\usepackage[bulletsep]{collref}
\let\oldbfseries=\bfseries
\let\oldmdseries=\mdseries
\let\oldnormalfont=\normalfont
\renewcommand{\bfseries}{\oldbfseries\boldmath}
\renewcommand{\mdseries}{\oldmdseries\unboldmath}
\renewcommand{\normalfont}{\oldnormalfont\unboldmath}

\allowdisplaybreaks[3]

\numberwithin{equation}{section}

\usepackage[font=small,labelfont=bf,width=0.85\textwidth]{caption}

\providecommand{\hypersetup}[1]{}
\providecommand{\texorpdfstring}[2]{#1}

\hypersetup{plainpages=false}
\hypersetup{pdfpagemode=UseNone}
\hypersetup{bookmarksnumbered=true}
\hypersetup{pdfstartview=FitH}
\hypersetup{colorlinks=false}
\hypersetup{citebordercolor={.5 1 .5}}
\hypersetup{urlbordercolor={.5 1 1}}
\hypersetup{linkbordercolor={1 .7 .7}}
\providecommand{\href}[2]{#2}
\providecommand{\arxivlink}[1]{\href{http://arxiv.org/abs/#1}{arxiv:#1}}

\newcommand{\eps}{\epsilon}

\newcommand{\Aa}{{\cal A}}
\newcommand{\FF}{{\cal F}}

\newcommand{\CC}{{\cal C}}

\newcommand{\OO}{{\cal O}}

\newcommand{\NN}{{\cal N}}
\newcommand{\TT}{{\cal T}}

\newcommand{\DD}{{\cal D}}

\newcommand{\VV}{{\cal V}}
\newcommand{\p}{{\partial}}
\newcommand{\wt}{\widetilde}

\newcommand{\up}{\uparrow}
\newcommand{\down}{\downarrow}

\newcommand{\ben}{\begin{eqnarray}\displaystyle}
\newcommand{\een}{\end{eqnarray}}

\newcommand{\al}{\alpha}

\newcommand{\s}{\sigma}

\newcommand{\la}{\lambda}

\newcommand{\Tr}{\hbox{Tr}}

\newcommand{\ve}{\varepsilon}
\newcommand{\De}{\Delta}
\newcommand{\de}{\delta}
\newcommand{\pdx}{\frac{\partial}{\partial x}}
\newcommand{\tM}{\widetilde M}
\newcommand{\tQ}{\widetilde Q}
\newcommand{\tS}{\widetilde S}
\newcommand{\dda}{{\dot{\alpha}}}
\newcommand{\ddb}{{\dot \beta}}
\newcommand{\ddg}{{\dot \gamma}}

\newcommand{\gym}{g_{\rm YM}}

\newcommand{\half}{\frac{1}{2}}

\begin{document}

%%%%%%%%%%%%%%%%%%%%%%%%%%%%%%%%%%%%%%%%%%%%%%%%%%%%%%%%%%%%%%%%%%%%%%%%%%%%%%%%
%%%%%%%%%%%%%%%%%%%%%%%%%%%%%%%%%%%%%%%%%%%%%%%%%%%%%%%%%%%%%%%%%%%%%%%%%%%%%%%%
% TITLE PAGE
\thispagestyle{empty}
\phantomsection
\addcontentsline{toc}{section}{Title}

\begin{flushright}\footnotesize%
\texttt{UUITP-38/10},
\texttt{\arxivlink{1012.3983}}\\
overview article: \texttt{\arxivlink{1012.3982}}%
\vspace{1em}%
\end{flushright}

%\thispagestyle{empty}
%%\phantomsection
%\addcontentsline{toc}{section}{Title}
%
%\begin{flushright}\footnotesize%
%\texttt{\arxivlink{yymm.nnnn}},
%\texttt{UUITP-38/10}
%\vspace{1em}%
%\end{flushright}

\begingroup\parindent0pt
\begingroup\bfseries\ifarxiv\Large\else\LARGE\fi
\hypersetup{pdftitle={Review of AdS/CFT Integrability, Chapter I.1: Spin Chains in N=4 Super Yang-Mills}}%
Review of AdS/CFT Integrability, Chapter I.1:\\
Spin Chains in $\NN=4$ Super Yang-Mills
\par\endgroup
\vspace{1.5em}
\begingroup\ifarxiv\scshape\else\large\fi%
\hypersetup{pdfauthor={Joseph A. Minahan}}%
Joseph~A.~Minahan
\par\endgroup
\vspace{1em}
\begingroup\itshape
Department of Physics and Astronomy, Uppsala University \\
SE-751 20 Uppsala, Sweden
\par\endgroup
\vspace{1em}
\begingroup\ttfamily
joseph.minahan@fysast.uu.se
\par\endgroup
\vspace{1.0em}
\endgroup

\begin{center}
\includegraphics[width=5cm]{TitleI1.mps}%figure for your chapter
\vspace{1.0em}
\end{center}

\paragraph{Abstract:}
In this chapter of {\it Review of AdS/CFT Integrability} we introduce $\NN=4$ Super Yang-Mills.  We discuss the global superalagebra $PSU(2,2|4)$ and its action on gauge invariant operators.  We then discuss the computation of the correlators of certain gauge invariant operators, the so-called single trace operators in the large $N$ limit.  We show that interactions in the gauge theory lead to mixing of the operators.  We compute this mixing at the one-loop level and show that the problem maps to a one-dimensional spin chain with nearest neighbor interactions.  For operators in the $SU(2)$ sector we show that the spin chain is the ferromagnetic Heisenberg spin chain whose eigenvalues are determined by the Bethe equations.

\ifarxiv\else
% http://www.ams.org/msc
\fi
\hypersetup{pdfsubject={MSC (2010): 81T60, 17A70, 82B23}}%

\ifarxiv\else
\fi
\hypersetup{pdfkeywords={N=4 SYM, spin-chains, Bethe equations}}%

\newpage

%%%%%%%%%%%%%%%%%%%%%%%%%%%%%%%%%%%%%%%%%%%%%%%%%%%%%%%%%%%%%%%%%%%%%%%%%%%%%%%%
%%%%%%%%%%%%%%%%%%%%%%%%%%%%%%%%%%%%%%%%%%%%%%%%%%%%%%%%%%%%%%%%%%%%%%%%%%%%%%%%
% BODY

%%%%%%%%%%%%%%%%%%%%%%%%%%%%%%%%%%%%%%%%%%%%%%%%%%%%%%%%%%%%%%%%%%%%%%%%%%%%%%%%
\section{Introduction and summary}

%\cite{Minahan:2002ve}
%see chapter \cite{ChapChain}
% It is possible to go beyond $\NN=4$ supersymmetry, but this necessarily introduces gravity and will not be considered here.  \dots \dots

In this chapter of {\it  Review of AdS/CFT Integrability} \cite{chapIntro}, we introduce $\NN=4$ super Yang-Mills (SYM),  a gauge theory with the maximal amount of  supersymmetry \footnote{This chapter is a substantial extension of an earlier review \cite{Minahan:2006sk}.}.  $\NN=4$ SYM was first considered by Brink, Scherk and Schwarz \cite{Brink:1976bc}, who explicity constructed its Lagrangian  by dimensionally reducing SYM from 10  to 4 dimensions.  One of the remarkable properties of $\NN=4$ SYM is that it is conformal \cite{Sohnius:1981sn}, meaning that it has no inherent mass scale in the theory.  Many theories are classically conformal, namely any theory with only massless fields and marginal couplings.  But $\NN=4$ stays conformal even at the quantum level.  In particular its $\beta$-function is zero to all orders in perturbation theory, as was first conjectured in \cite{Green:1982sw} when studying open string loop amplitudes which reduce to ten dimensional SYM in the infinite string tension limit.  

In a theory such as QCD  which has a running coupling constant, there is a natural mass   scale at the crossover point from weak to strong coupling.  In QCD this is roughly where confinement sets in and is responsible for the   proton mass.
Since $\NN=4$ SYM is conformal it cannot be confining, meaning that there are no mesons and hadrons, the physical particles in QCD.  Why then should we study it?  

There are several reasons.  First, its large amount of symmetry leads to an underlying integrability, making many physical quantities analytically calculable, as many of the chapters in this review will explain.  Second, the AdS/CFT correspondence \cite{Maldacena:1997re,Gubser:1998bc,Witten:1998qj} conjectures that $\NN=4$ Super Yang-Mills is equivalent to type IIB string theory on $AdS_5\times S^5$.  This correspondence is a strong/weak duality which is normally very difficult to confirm because when one theory is computationally  under control the other is not.   However,  the integrability allows us to plow forward and calculate at strong coupling, thus testing many consequences of the conjecture.  Third, while QCD is not conformal, it is asymptotically free.  Hence at high energies it is close to being conformal.    Many essential features of high energy gluon scattering, which is relevant for the LHC, can be learned by studying gauge boson amplitudes in $\NN=4$ SYM.  

There are other reasons for studying $\NN=4$ SYM, including its conjectured invariance under $SL(2,Z)$ duality transformations \cite{Montonen:1977sn,Osborn:1979tq,Girardello:1995gf,Vafa:1994tf}, but they are less relevant for integrability.  Nevertheless, the three reasons stated here are hopefully enough motivation to press on.

In the following sections 
 we will first describe the fields that make up $\NN=4$ SYM, showing that they lead to a vanishing one-loop $\beta$-function.   
We then discuss the symmetry algebra of $\NN=4$.  Here we define a class of operators called chiral primaries whose dimensions are protected from quantum corrections.     We next describe a particular set of gauge invariant operators, single trace operators, which are of significant importance in the large $N$ limit.  We find how the fields transform under the symmetry algebra and from there find the chiral primaries in the  single trace operators.    Using supersymmetry arguments we then show that the gauge coupling $\gym$ is fixed under rescalings and so the theory is conformal, even at the quantum level.  

We then compute the one-loop anomalous dimensions for a general set of single trace operators composed of scalar fields.  We show that in the large $N$ limit where the contributions to the anomalous dimensions are dominated by planar graphs, the problem is identical to computing the energies of a certain spin-chain with nearest neighbor interactions.  We then describe how the spin-chain can be  generalized to all single trace operators.  Finally, we discuss the solutions for this spin-chain in a particular sector called the $SU(2)$ sector, where one finds the famous Bethe equations.

The full description of these spin chains, including their higher loop generalizations  and their solutions are deferred to later chapters of the review.

\section{The field content and the vanishing \texorpdfstring{$\beta$}{beta}-function}

The fields  contained in $\NN=4$ SYM are the gauge bosons $\Aa_\mu$,  six massless real scalar fields $\phi^I$, $I=1\dots 6$, four chiral fermions $\psi^a_{\al}$ and four anti-chiral fermions $\overline\psi_{\dot\al\,a}$, with $a=1\dots 4$.   The indices   $\al,\dda=1,2$  are the spinor indices  of the two independent $SU(2)$ algebras that make up the 4 dimensional Lorentz algebra.  All fields transform  in the adjoint representation of the $SU(N)$ gauge group.  There is a global $SU(4)\simeq SO(6)$ symmetry, called an $R$-symmetry, with the scalars transforming in the  {\bf 6}, $\psi^a_{\al}$ in the {\bf 4} (raised $a$ index) and $\overline\psi_{\dot\al\, a }$ in the $\bf\overline4$  (lowered  $a$ index) representations of the $R$-symmetry algebra.   

Let us use the information about the field content to rapidly show that the one-loop $\beta$-function is zero.  For any $SU(N)$ gauge theory, the one-loop $\beta$-function for the gauge coupling $\gym$ is given by \cite{Gross:1973ju}
\begin{equation}
\beta_1(\gym)\equiv \mu\frac{\partial \gym}{\partial\mu}=-\frac{\gym^3}{16\pi^2}\left(\frac{11}{3}\,N-\frac{1}{6}\sum_{i} C_i-\frac{1}{3}\sum_j \widetilde C_j\right)\,,
\end{equation}
where the first sum is over all real scalars with quadratic casimir $C_i$ and the second sum is over all Weyl fermions with quadratic casimir $\widetilde C_j$.  All fields in $\NN=4$ SYM are in the adjoint, hence all casimirs are $N$.  One can then quickly see that with six real scalars and eight Weyl fermions that $\beta_1(\gym)=0$.

Going beyond one-loop, the $\beta$-function for $\NN=4$ SYM was shown to be zero up to three loops using superspace arguments \cite{Grisaru:1980nk}.  Subsequently it was argued using light cone gauge that the $\beta$-function is zero to all loops \cite{Mandelstam:1982cb,Brink:1982wv}.  In a later section we will present a different  argument for why the $\beta$-function is zero to all orders.

\section{The superconformal algebra}

The conformal symmetry,  the supersymmetery and the $R$-symmetry of $\NN=4$ SYM are part of a larger symmetry group.  This group is known as the $\NN=4$ superconformal group, or more formally as $PSU(2,2|4)$.    This symmetry group is unbroken by quantum corrections and thus serves as a powerful tool by putting significant constraints on the theory.  In this section we will review the $PSU(2,2|4)$ algebra and its consequences.  A more detailed description is given in \cite{chapSuperconf}.

$PSU(2,2|4)$ has the bosonic subalgebra $SU(2,2)\times SU(4)$. The $SU(2,2)\simeq SO(2,4)$ is the four dimensional conformal algebra while the $SU(4)\simeq SO(6)$ is the $R$-symmetry.  The conformal algebra has 15 generators: ten generators belong to the Poincar\'e algebra which itself contains four generators of space-time translations, $P_\mu$ and six generators of the $SO(1,3)\equiv SU(2)\times SU(2)$ Lorentz transformations, $M_{\mu\nu}$.  The other generators of the conformal algebra are the four generators of special conformal transformations, $K_\mu$ and  one generator of dilatations,
$D$.   These generators then satisfy the commutation relations
\begin{eqnarray}\label{confalg}
&&\qquad[D,P_\mu]=-iP_\mu\qquad[D,M_{\mu\nu}]=0\qquad[D,K_\mu]=+iK_\mu\nonumber\\
&&[M_{\mu\nu},P_\lambda]=-i(\eta_{\mu\la}P_\nu-\eta_{\la\nu}P_{\mu})\qquad
 [M_{\mu\nu},K_\lambda]=-i(\eta_{\mu\la}K_\nu-\eta_{\la\nu}K_{\mu})\qquad\nonumber\\
&&\qquad\qquad\qquad[P_\mu,K_\nu]=2i(M_{\mu\nu}-\eta_{\mu\nu}D)\,.
 \end{eqnarray}

Let $\OO(x)$ be  a local operator in the field theory with  dimension $\De$.  This signifies that under the rescaling $x\to \la x$, $\OO(x)$ scales as  $\OO(x)\to \la^{-\De}\OO(\la x)$.   $D$ is the generator of these scalings, by which we mean that $\OO(x)\to\la^{-i\,D}\OO(x)\la^{i\, D}$. Thus, its action on $\OO(x)$ is 
\begin{eqnarray}
[D,\OO(x)]=i\left(-\De+x\,\pdx\right)\OO(x)\,.
\end{eqnarray}
Next, we let $D$ act on $[K_\mu,\OO(0)]$, where we find using the Jacobi identity
\begin{eqnarray}
[D,[K_\mu,\OO(0)]]&=&[[D,K_\mu],\OO(0)]+[K_\mu,[D,\OO(0)]\nonumber\\
&=&i[K_\mu,\OO(0)]-i\De[K_\mu,\OO(0)]\,.
\end{eqnarray}
Thus, $K_\mu$ creates a new local operator from  $\OO$ with its dimension lowered by 1.  Aside from the identity operator, the local operators  in a unitary quantum field theory must have positive dimension.  Therefore, if we keep creating new lower dimensional operators by commuting with the special conformal generators, we must eventually reach a barrier where we can go no further.  Hence the last operator in this chain, $\wt\OO(x)$ must satisfy
\begin{eqnarray}\label{primcond}
[K_\mu,\wt\OO(0)]=0\,.
\end{eqnarray}
for all $K_\mu$.
The  operator $\wt\OO(x)$ is called {\it primary}\footnote{The primary condition (\ref{primcond}) is defined at $x=0$ where the space-time position is a fixed point of the dilatation.  If the local operator were at a different space-time point then it would commute with a different combination of the conformal generators.}.  Starting with $\wt\OO$, we can build new operators with the same dimension or higher by commuting it with the other generators of the conformal algebra.  The higher dimensional  operators are called {\it descendants}\footnote{Peradventure they should have been called {\it ascendants}.} of $\wt\OO$.

The conformal algebra can be combined with supersymmetry to make a superconformal algebra.  In four  dimensions one can have gauge theories with $\NN=1$, $\NN=2$ or $\NN=4$ supersymmetry, and all of these cases can be combined with the conformal symmetries to make an $\NN=1$, $\NN=2$ or $\NN=4$ superconformal algebra.  Here, we only consider the $\NN=4$ case.  

The generators of   supersymmetry transformations are fermionic and are called {\it supercharges}.  For $\NN=4$ supersymmetry there are 16 separate supercharges, $Q_{\al\,a}$ and $\tQ^a_{\dda}$, where $\al,\dda=1,2$ and  $a=1..4$ are the same spinor and $R$-symmmetry indices that label the Weyl fields, except here the $\al$ indices are paired with the $\bf\overline4$ and the $\dot\al$ indices are paired with the ${\bf 4}$.  The supersymmetry algebra is   a {graded} Lie algebra which combines the generators of the Poincar\'e algebra with the supercharges and contains  the commutation and anti-commutation relations
\begin{eqnarray}\label{QQ}
\{Q_{\al\,a},\tQ^b_{\dda}\}&=&\gamma^\mu_{\al\dda}{\de_{a}}^bP_\mu\,,\qquad\{Q_{\al\,a},Q_{\al\,b}\}=
\{\tQ^a_{\dda},\tQ^b_{\dda}\}=0\nonumber\\
&&[ P_\mu, Q_{\al\,a}]=[P_\mu,\tQ^b_{\dda}]=0\nonumber\\
&&[M^{\mu\nu},Q_{\al\,a}]=i\gamma^{\mu\nu}_{\al\beta}\eps^{\beta\gamma}Q_{\gamma\,a}\,,\quad[M^{\mu\nu},\tQ^a_{\dda}]=i\gamma^{\mu\nu}_{\dda\ddb}\eps^{\ddb\ddg}\tQ^a_{\ddg}\,,
\end{eqnarray}
where  $\gamma^{\mu\nu}_{\al\beta}=\gamma^{[\mu}_{\al\dda}\gamma^{\nu]}_{\beta\ddb}\ve^{\dda\ddb}$.
  Simple dimension counting within the algebra shows that  $Q_{\al\,a}$ and $\tQ^a_{\dda}$ have dimension $1/2$ and so their commutators with $D$ is
\begin{eqnarray}
   [D,Q_{\al\,a}]=-\frac{i}{2}Q_{\al\,a}\qquad\qquad[D,\tQ^a_{\dda}]=-\frac{i}{2}\tQ^a_{\dda}\,.
 \end{eqnarray} 

By including the special conformal generators we generate a new set of  supercharges by  commuting $K_\mu$ with $Q_{\al\,a}$ and $\tQ^a_{\dda}$, 
\begin{eqnarray}
[K^\mu,Q_{\al\,a}]=\gamma^\mu_{\al\dda}\eps^{\dda\ddb}\tS_{\ddb\,a}\qquad
[K^\mu, \tQ^a_{\dda}]=\gamma^\mu_{\al\dda}\eps^{\al\beta}S^a_{\beta}\,.
\end{eqnarray}
The operators $S^a_{\al} $ and $\tS_{\dda\,a}$ have dimension $-1/2$ and are known as the special conformal supercharges, or the superconformal charges.     Their $R$-charge representations are reversed from the supercharges and combine with the regular supercharges to give 32 supercharges in total.
The superconformal generators have  anticommutation relations that mirror the anticommutation relations of the supercharges,
\begin{eqnarray}\label{scsccr}
\{S^a_{\al},\tS_{\dda\,b}\}&=&\gamma^\mu_{\al\dda}{\de^a}_bK_\mu\qquad\{S^a_{\al},S^b_{\al}\}=
\{\tS_{\dda\,a},\tS_{\dda\,b}\}=0\,\nonumber\\
{} [K_\mu,S^a_{\al}]&=&[K_\mu,\tS_{\dda\,a}]=0\,.
\end{eqnarray}
Nonzero anticommutation relations between  the supercharges and the superconformal charges complete the algebra,
\begin{eqnarray}\label{QS}
\{Q_{\al\,a},S^b_{\beta}\}&=&-i\varepsilon_{\al\beta}{{\s^{IJ}}_a}^ bR_{IJ}+\gamma^{\mu\nu}_{\al\beta}{\de_a}^bM_{\mu\nu}-\frac{1}{2}\varepsilon_{\al\beta}{\de_a}^b\,D\nonumber\\
\{ \tQ^a_{\dda},\tS_{\ddb\,b}\}&=&+i\varepsilon_{\dda\ddb}{\s^{IJ\,a}}_{b}\,R_{IJ}+\gamma^{\mu\nu}_{\dda\ddb}{\de^a}_bM_{\mu\nu}-\frac{1}{2}\varepsilon_{\dda\ddb}{\de^a}_bD\nonumber\\
\{Q_{\al\,a},\tS_{\ddb\,b}\}&=&\{\tQ^a_{\dda},S^b_{\beta}\}=0\,.
\end{eqnarray}
On the righthand side of (\ref{QS}) one has in addition to  the Lorentz and dilatation generators  the $SU(4)\simeq SO(6)$ $R$-symmetry generators $R_{IJ}$, where $I,J=1\dots6$.    The supercharges transform under the two spinor representations of  $SO(6)$, while all generators of the conformal algebra commute with $R_{IJ}$.   

Let us now return to the primary operator $\wt\OO(x)$.  Commuting the superconformal charges with a local operator $\OO(0)$ lowers the dimension by 1/2.  A lower bound on the dimension must still exist, so we assume that $\wt\OO(0)$ satisfies  
\begin{equation}
[S^a_{\al},\wt\OO(0)]=[\tS_{\dda\,a},\wt\OO(0)]=0\qquad\qquad{\rm for\ all\ } \al,\dda,a\,.
\end{equation}
$\wt\OO(x)$ is clearly primary since  the anticommutation relations in (\ref{scsccr}) directly lead to (\ref{primcond}).  The descendants of $\wt\OO(0)$ are constructed from the rest of the algebra.

The primary operator and its descendants make up an irreducible representation of $PSU(2,2|4)$, with the primary as  the highest weight of the representation.   $PSU(2,2|4)$ is noncompact, so the representation is infinite dimensional\footnote{Except for the trivial representation which only contains the identity operator.}.  For example, one can act with $P_\mu$ on $\wt\OO(x)$ an arbitrary number of times, where $[P_\mu,\OO(x)]=-i\partial_\mu\OO(x)$, making a new local operator with one higher dimension.    Using the supercharges we can also make new operators with $1/2$ higher dimension.

We will be particularly interested in a class of highest weight representations which, while still infinite dimensional, are smaller because  there are fewer independent operators at each half-step in dimension.  In order for this to occur, $\wt\OO(0)$ must commute with some of the supercharges.  Let us then place the further restriction on $\wt\OO(x)$ that 
\begin{equation}
[Q_\al^a,\wt\OO(0)]=0\qquad\qquad{\rm for\ some\ } \al,a\,.
\end{equation}
It then follows from the anticommutation relations in (\ref{QS}) that
\begin{eqnarray}
[\{Q_{\al\,a},S^b_{\beta}\},\wt\OO(0)]&=&%\{Q_\al^a,[S^\bab_\beta,\OO(0)]\}+\{S^\bab_\beta,[Q_\al^a,\OO(0)]\}=0\nonumber\\
[-i\ve_{\al\beta}{{\s^{IJ}}_a}^bR_{IJ}-\ve_{\al\beta}{\de_{a}}^b\,D+\s^{\mu\nu}_{\al\beta}{\de_{a}}^bM_{\mu\nu},\wt\OO(0)]=0\,.\nonumber\\
\end{eqnarray}
We assume that  $\wt\OO(x)$ is a scalar,  therefore $\wt\OO(0)$ commutes with the Lorentz generators  $M_{\mu\nu}$.   What remains is a simple relation between the action of the $R$-symmetry and the dimension $\Delta$ of $\wt\OO(x)$,
\begin{equation}\label{chprrel}
{{\s^{IJ}}_a}^b[R_{IJ},\wt\OO(0)]=\Delta\,{\de_{a}}^b\ \wt\OO(0)\,.
\end{equation}

To help us find operators that can satisfy the relation in (\ref{chprrel}) we consider the Cartan subalgebra of $SO(6)$.  $SO(6)$ is a rank 3 group and thus has three commuting generators in its Cartan subalgebra. We choose these generators to be $R_{12}$, $R_{34}$ and $R_{56}$ 
and  write the corresponding charges  as $(J_1,J_2,J_3)$.  The ${\s^{IJ\,a}}_{b}$ are the  generators in the $SU(4)$ fundamental representation, with 
\begin{equation}
{\s^{12}}={\scriptsize\left(\begin{array}{cccc}1&0&0&0\\
                                          0&1&0&0\\
                                          0&0&-1&0\\
                                          0&0&0&-1\end{array}\right)}\quad
\s^{34}={\scriptsize\left(\begin{array}{cccc}1&0&0&0\\
                                          0&-1&0&0\\
                                          0&0&1&0\\
                                          0&0&0&-1\end{array}\right)}\quad                                          
 \s^{56}={\scriptsize\left(\begin{array}{cccc}1&0&0&0\\
                                          0&-1&0&0\\
                                          0&0&-1&0\\
                                          0&0&0&1\end{array}\right)}\,,
\end{equation}
as a consistent choice of Cartan generators.
Hence, a primary operator with $R$-charges $(J_1,0,0)$ is  annihilated by $Q_{\al1}$ and $Q_{\al2}$ if $\Delta=J_1$.  The anticommutation relations in  (\ref{QS}) indicate that such operators are also annihilated by $\tQ^{\,3}_{\dot\al}$ and $\tQ^{\,4}_{\dot\al}$.  Hence, an operator of this type commutes with half of the supercharges.  Such operators are called {chiral primary} or BPS operators.  By the same logic an operator with $(0,J_2,0)$ and dimension $\Delta=J_2$ is also a chiral primary.  But such a state is in the same $SO(6)$ representation as the $(J_2,0,0)$ operator, and hence is in the same $PSU(2,2|4)$ representation.  Therefore, it is only necessary to consider the scalar operators with charges $(J,0,0)$ and $\Delta=J$. 

In general the dimension of an operator will depend on the Yang-Mills coupling $\gym$.  The dimension at zero coupling is known as the bare dimension.  The correction to the bare dimension is the anomalous dimension.  From our discussion so far we learn two important facts.  First, the anomalous dimensions within the same $PSU(2,2|4)$ representation are equal.  This is because the generators can only change the dimension in $1/2$ integer steps.  Second, and more strikingly, the chiral primaries and their descendants cannot have an anomalous dimension.  This is because the chiral primaries commute with half the supercharges no matter what the coupling.  If they did not commute then there would have to be extra operators at each level.  But the number of independent operators with a given dimension is a finite integer which cannot change by varying a continuous parameter such as the coupling.  Hence, the relation in (\ref{chprrel}) continues to hold.  Since the $R$-charges are integers  that stay fixed, then  the dimensions must also stay fixed.

\section{Gauge invariant operators in \texorpdfstring{$\NN=4$}{N=4} SYM}

We now apply our discussion in the previous section to the actual operators that one encounters in $\NN=4$ SYM.  The physical observables in a gauge theory must be gauge invariant.  In $\NN=4$ SYM, the local gauge invariant operators are made up of products of traces of  the fields that transform covariantly under the gauge group.  This includes  
the scalars $\phi^I$, the fermions $\psi^a_\al$, $\overline\psi_{\dda\,a}$ and the field strengths $\FF_{\mu\nu}$.  Since these fields all lie in the adjoint representation, their transformation under a gauge transformation is\begin{equation}\label{gt}
\chi(x)\to\chi(x)+[\ve(x) ,\chi(x)]
\end{equation}
where $\chi(x)$ is one of the covariant fields and $\ve(x)$ is a generator of gauge transformations. We have explicitly included the space-time dependence of the fields to emphasize that this is a local transformation.  %
From a covariant field $\chi(x)$ we can make other covariant fields $\DD_\mu\chi(x)$, where $\DD_\mu$ is the covariant derivative
\begin{equation}
\DD_\mu\chi(x)\equiv\p_\mu\chi(x)-[\Aa_\mu(x),\chi(x)]\,.
\end{equation}
The gauge connection $\Aa_\mu(x)$ does not transform covariantly, but instead transforms as
\begin{equation}
\Aa_\mu(x)\to\Aa_\mu(x)+\p_\mu\ve(x)+[\ve(x),\Aa_\mu(x)].
\end{equation}

It is then clear that the single trace local operator
\begin{equation}
\OO(x)=\Tr[\chi_1(x)\chi_2(x)...\chi_L(x)]\,,
\end{equation}
where $\chi_i(x)$ refers to one of the above covariant fields with or without covariant derivatives, is gauge invariant.  
We can also build other local gauge invariant operators by taking products of traces.  
Later on we will take the limit where  the number of colors $N$ is large. 
In this  limit the dimension of the product of single trace operators is equal to the sum of their dimensions, so all information about the spectrum of local operators comes from the single trace operators.

Because $[\DD_\mu,\DD_\nu]=-\FF_{\mu\nu}(x)$, any antisymmetric combination of covariant derivatives can always be replaced with a field strength.  Hence, it is only necessary to consider symmetric products of $\DD_\mu$ acting on any field $\chi$.  Furthermore, we can use the equations of motion and the Bianchi identities  to get rid of certain combinations of covariant derivatives.    As an example, the equations of motion for the scalar fields are schematically
\begin{equation}\label{sceom}
\DD^\mu\DD_\mu \phi^I=...\,.
\end{equation}
The right hand side of (\ref{sceom}) contains cubic scalar terms as well as fermion bilinears, but otherwise has no derivatives.  Therefore,  inside a trace we can always replace two contracted derivatives on a scalar with nonderivative terms.

With these rules we can build all single trace operators.
We first construct the single trace chiral primaries,  from which we can systematically assemble the other   operators.  
The $SU(2,2)\times SU(4)$ bosonic subgroup of  $PSU(2,2|4)$ is rank six and so an operator  will have a sextuplet of charges,  $(\Delta,S_1,S_2;J_1,J_2,J_3)$.  The $J_i$ are the $R$-charges discussed in the last section,  $\Delta$ is the dimension, and $S_1$ and $S_2$ are the two charges of the $SO(1,3)$ Lorentz group ({\it i.e.} the spins).  In this subsection we will only consider the gauge theory at zero coupling, in which case the dimension can be replaced with the bare dimension $\Delta_0$ and all dimensions are additive. 

The six adjoint scalars $\phi^I$ can be expressed as three complex fields, $Z=\frac{1}{\sqrt{2}}(\phi^1+i\phi^2)$, $W=\frac{1}{\sqrt{2}}(\phi^3+i\phi^4)$ $X=\frac{1}{\sqrt{2}}(\phi^5+i\phi^6)$, along with their conjugates.  Scalars in 4 dimensions have bare dimension 1 and are of course spinless, thus
the charges for $Z$, $W$ and $X$  are given by $(1,0,0;1,0,0)$, $(1,0,0;0,1,0)$,  and $(1,0,0;0,0,1)$ respectively.  Their conjugates, $\bar Z$, $\bar W$ and $\bar X$ have reversed $R$-charges.  The sixteen fermions $\psi_\alpha^a$ and $\overline\psi_{\dot\alpha\, a}$ have charges $(\frac32,\pm \frac12,0;\pm\frac12,\pm\frac12,\pm\frac12)$ and $(\frac32,0,\pm \frac12;\pm\frac12,\pm\frac12,\pm\frac12)$ where the number of  negative signs for the $SU(4)$ charges is even for the first set and odd for the second.  The field strengths have six independent components and naturally split into their even and odd self-duals, where $\FF_{\pm}^{\mu\nu}=\pm \frac12\varepsilon^{\mu\nu\s\rho}\FF_{\pm\mu\nu}$.  In terms of the $SO(1,3)\simeq SU(2)\times SU(2)$ Lorentz group, the even and odd self-duals fall into the $(3,1)\oplus(1,3)$ representation.  It is thus convenient to write the components using the $ SU(2)\times SU(2)$ spinor indices, where we define 
\begin{equation}
\FF_{+\al\beta}\equiv\half( \gamma^{\mu\nu})_{\al\beta}\FF_{+\mu\nu}=\FF_{+\beta\al}\,,
\qquad
\FF_{-\dot\al\dot\beta}\equiv\half (\gamma^{\mu\nu})_{\dot\al\dot\beta}\FF_{-\mu\nu}=\FF_{-\dot\beta\dot\al}\,.
\end{equation}
From this we readily see that the $\FF_+$  have charges $(2,m,0;0,0,0)$ and the $\FF_-$ have charges $(2,0,m;0,0,0)$ where $m=+1,0,-1$.  It is also useful to write the covariant derivatives as a bispinor $\DD_{\al\dot\beta}\equiv(\gamma^\mu)_{\al\dot\beta}\DD_\mu$.   Then $D_{\al\ddb}$ acting on a field adds the charges $(1,\pm\half,\pm\half;0,0,0)$ to the charges of the operator.

Let us now consider the gauge invariant operator $\Psi_L\equiv\Tr[Z^L]$, with $L\ge2$ ($\Tr Z=0$).  The charges of $\Psi_L$ are $(L,0,0;L,0,0)$, which satisfies $\Delta_0=J_1$.  Therefore, $\Psi_L$  is a chiral primary and $\Delta=\Delta_0$, even after the coupling is turned on. $\Psi_L$ is the highest weight element of the $L$-fold symmetric traceless representation of $SO(6)$.  Hence, any operator of the form
$$\chi_{I_1I_2\dots I_L}\Tr(\phi^{I_1}\phi^{I_2}\dots\phi^{I_L})\,,$$
where $\chi_{I_1I_2\dots I_L}$ is completely symmetric in its indices and the trace of any two indices is zero, is a chiral primary with its dimension protected from quantum corrections. Notice further that  if we change one of the $Z$ fields in $\Psi_L$  to any other scalar field, aside from $\overline Z$, then the resulting operator  it is automatically symmetric and traceless because of the cyclicity of the trace.  To make a non-BPS operator strictly out of scalars will  require at least one $\overline Z$ or two other scalar fields that are not  $Z$ or $\overline Z$.

A very convenient way to classify the single trace operators is to use bosonic and fermionic creation operators \cite{Gunaydin:1981yq,Gunaydin:1984fk,Gunaydin:1998sw,Gunaydin:1998jc} (see also \cite{chapSuperconf}).  To this end we note that the vector representation of  the $SO(6)$ $R$-symmetry group is equivalent to the antisymmetric representation of $SU(4)$.  Hence, the scalar fields can be written in $SU(4)$ notation as $\phi^{ab}$ with the indices antisymmetrized.  Likewise, the fermions in the antifundamental representation can have its $SU(4)$ index raised to three antisymmetric indices, $\overline\psi^{abc}_{\dot\al}\equiv\ve^{abcd}\overline\psi_{\dot\al\,d}$.  Thus all fields have their fundamental  $SU(4)$ indices  antisymmetrized.  Furthermore, the field strengths come with symmetrized spinor indices, the combination $\ve^{\al\beta}\DD_{\al\dda}\psi_\beta^a$ can always be replaced by a nonderivative term by the equations of motion, and all covariant derivatives are symmetrized.  Hence, all indices in either of the  $SU(2)$'s  of the Lorentz group are symmetrized for any field, including those with covariant derivatives.  

Therefore, we will build the fields at each site within the trace with two sets of bosonic creation operators $A^\dag_\alpha$, $B^\dag_{\dda}$,  and a set of fermionic creation operators $C^{a\dag}$.   The adjoints of these fields are $A^\al$, $B^\dda$ and $C_a$ and we have the usual set of commutation or anticommutation relations
\begin{equation}
[A^\al,A^\dag_\beta]={\delta^\al}_\beta\,,\qquad[B^\dda,B^\dag_{\ddb}]={\delta^\dda}_\ddb\,,\qquad\{C_a,C^{b\dag}\}={\delta_a}^b\,.
\end{equation}
One starts with a ground state $|0\rangle$ for each site and defines the operator
\begin{equation}
\CC=A^\dag_\al A^\al-B^\dag_\dda B^\dda+C^{a\dag}C_a-2\,. 
\end{equation}
 Then the states that correspond to the actual fields are those states $|\chi\rangle$ in the oscillator Fock space  where $\CC|\chi\rangle=0$.  We denote this projected Fock space by $\VV$.  The states satisfying the $\CC=0$ condition and the fields they correspond to are
\begin{eqnarray} \label{singleton}
(A^\dag)^{k+2}(B^\dag)^k|0\rangle\ \ \ \Rightarrow&&\ \DD^k\FF_+\nonumber\\
(A^\dag)^{k+1}(B^\dag)^kC^{a\dag}|0\rangle\ \ \ \Rightarrow&&\ \DD^k\psi^{a}\nonumber\\
(A^\dag)^{k}(B^\dag)^kC^{a\dag}C^{b\dag}|0\rangle\ \ \ \Rightarrow&&\ \DD^k\phi^{ab}\nonumber\\
(A^\dag)^{k}(B^\dag)^{k+1}C^{a\dag}C^{b\dag}C^{c\dag}|0\rangle\ \ \ \Rightarrow&&\ \DD^k\overline\psi^{abc}\nonumber\\ 
(A^\dag)^{k}(B^\dag)^{k+2}C^{a\dag}C^{b\dag}C^{c\dag}C^{d\dag}|0\rangle\ \ \ \Rightarrow&&\ \DD^k\FF_-\,,
\end{eqnarray}
where we have suppressed all Lorentz indices.   

The elements of $PSU(2,2|4)$ can also be nicely represented by the oscillators.  In particular we have that
\begin{eqnarray}\label{algosc}
P_{\al\ddb}=A^\dag_\al B^\dag_\ddb&& K_{\al\ddb}=-\ve_{\al\gamma}\ve_{\ddb\dot\delta} A^\gamma B^{\dot\delta}\nonumber\\
Q_{\al\,a}=A^\dag_\al C_a\qquad \tQ^a_{\dda}=B^\dag_\dda C^{a\dag}&&\qquad S_\al^a=-i\ve_{\al\beta}A^\alpha C^{a\dag}\qquad \tS_{\dda\,a}=-i\ve_{\dda\ddb}B^\ddb C_a\nonumber\\
{R^a}_b=C^{a\dag}C_b-\frac{1}{4}{\delta^a}_b C^{c\dag}C_c&&D=-\frac{i}{2}\left(A^\dag_\al A^\al+B^\dag_\dda B^\dda+2\right)\nonumber\\
{M_\al}^\beta=A^\dag_\al A^\beta-\frac{1}{2}{\delta_\al}^\beta A^\dag_\gamma A^\gamma &&{\tM_\dda}^{\ \ddb}=B^\dag_\dda B^\ddb-\frac{1}{2}{\delta_\dda}^\ddb B^\dag_{\dot\gamma} B^{\dot\gamma}\,,
\end{eqnarray}
where we have expressed the $R$-symmetry generators in $SU(4)$ notation and the Lorentz generators in $SU(2)\times SU(2)$ notation \footnote{Strictly speaking, one should use the conformal Hamiltonian, $H=i\,D$, instead of $D$ as an $SU(2,2)$ generator.    See \cite{chapSuperconf} for a further discussion on this point.}. The oscillator representation of the algebra is also useful when applied to $\NN=$ SYM scattering amplitudes \cite{chapDual}. Notice that all  generators commute with $\CC$, hence $\CC$ is a centralizer of the algebra.  Thus, the elements of the algebra acting on the above states preserve the $\CC=0$ condition.  In fact the ``$P$" in front of $PSU(2,2|4)$ stands for ``projective" and corresponds to the projection we have made onto the $\CC=0$ states.  This projection is necessary  in order for  (\ref{algosc}) to give the relations in (\ref{QS}).  

The set of projected states in this Fock space (\ref{singleton}) form an irreducible representation of $PSU(2,2|4)$ called the ``singleton"  representation \cite{Ferrara:1998ej}\footnote{Some authors call this representation a ``doubleton" ({\it cf.} \cite{Gunaydin:1998sw}).  Another name is the fundamental representation.}.  However, it cannot correspond to a representation of gauge invariant operators since all of the fields are traceless.  Hence we will need $L\ge2$ fields inside the trace, leading  to tensor products of the singleton representations.  
\begin{equation}\label{tensorprod}
\VV_1\otimes\VV_2\otimes\dots\otimes\VV_L\,.
\end{equation}
The various generators of $PSU(2,2|4)$ on the tensor product have the general form
\begin{equation}\TT=\sum_{\ell=1}^L\oplus \TT_\ell\,,
\end{equation}
where $\TT_\ell$ is the generator at site $\ell$.  We can also define $\CC$ in this way, however the projection is still carried out at each site, {\it i.e.} $\CC_\ell=0$.  A gauge invariant operator is then mapped to a state in the tensor product, but because of the cyclicity of the trace must be projected onto only those states that are invariant under the shift, 
\begin{equation}\label{shift1}
\VV_1\otimes\VV_2\otimes\dots\otimes\VV_L\to\VV_L\otimes\VV_1\otimes\dots\otimes\VV_{L-1}\,.
\end{equation}

Let us now concentrate on the operator 
\begin{equation}
\OO^{abcd}=\Tr \phi^{ab}\phi^{cd}-\frac{1}{4!}\ve^{abcd}\ve_{a'b'c'd'}\Tr \phi^{a'b}\phi^{c'd'}\,,
\end{equation}
 which is part of the same $SU(4)$ representation as $\Tr Z^2$ and so is a chiral primary.  We then act with four supercharges in the following manner:
\begin{equation}
\frac{1}{4!}\ve^{\al\gamma}\ve^{\beta\delta}\{Q_{\al\,a},[Q_{\beta\,b},\{Q_{\gamma\,c},[Q_{\delta\,d},\OO^{abcd}]\}]\}=\ve^{\al\gamma}\ve^{\beta\delta}\Tr\FF_{+\al\beta}F_{+\gamma\delta}\,.
\end{equation}
Likewise, letting $\OO_{abcd}=\Tr \phi_{ab}\phi_{cd}-\frac{1}{4!}\ve_{abcd}\ve^{a'b'c'd'}\Tr \phi_{a'b}\phi_{c'd'}$ and acting with the other four supercharges we find
\begin{equation}
\frac{1}{4!}\ve^{\dot\al\dot\gamma}\ve^{\beta\delta}\{\tQ_{\dda}^a,[\tQ_{\ddb}^b,\{\tQ_{\dot\gamma}^c,[\tQ_{\dot\delta}^d,\OO_{abcd}]\}]\}=\ve^{\al\gamma}\ve^{\beta\delta}\Tr\FF_{-\al\beta}F_{-\gamma\delta}\,.
\end{equation}
Therefore, $\Tr\FF_{+}F_{+}$  and $\Tr\FF_-\FF_-$ are in the same supermultiplet as the chiral primary and thus their dimensions are protected from quantum corrections, meaning that they have dimension 4 no matter what the coupling.  These terms appear in the Lagrangian under the combination 
\begin{equation}
-i(\tau\,\Tr\FF_{+}F_{+}-\overline\tau\,\Tr\FF_-\FF_-)\,,
\end{equation}
where $\tau=\frac{4\pi i}{\gym^2}+\frac{\theta}{2\pi}$ with the $\theta$-angle included. Since  $\Tr\FF_{+}F_{+}$ and $\Tr\FF_-\FF_-$ are dimension 4 and the Lagrangian must also be dimension 4, we see that $\tau$, and hence $\gym$ is invariant under rescaling.  From this argument we learn that the $\beta$-function is zero.

\section{One loop anomalous dimensions and the relation to spin chains}

In this section we compute the one-loop anomalous dimensions for operators composed of scalar fields with no covariant derivatives \cite{Minahan:2002ve}.  This computation is complicated by the problem of operator mixing.  However, the mixing can often be restricted to operators within certain ``closed" sectors.  

To find  the anomalous dimension of an operator, one considers the two-point correlator of the operator with itself.  In particular, one finds that
\begin{equation}\label{2pt}
\langle \OO(x)\overline\OO(y)\rangle\approx \frac{1}{|x-y|^{2\Delta}}\,,
\end{equation}
where the dimension $\Delta=\Delta_0+\gamma$, with $\Delta_0$ being the bare dimension and $\gamma$ being the anomalous dimension arising from quantum corrections.  
For operators made up only of scalar fields with no covariant derivatives, all fields have bare dimension 1 and the bare dimension of the operator is $L$, the number of scalar  fields inside the trace.

If $\gym$ is small, then $\gamma<<\Delta_0$, in which case we can approximate the correlator in (\ref{2pt}) as
\begin{equation}\label{2pta}
\langle \OO(x)\overline\OO(y)\rangle\approx \frac{1}{|x-y|^{2\Delta_0}}(1-\gamma\,\ln\Lambda^2|x-y|^2)\,,
\end{equation}
where $\Lambda$ is cutoff scale.  The leading contribution to this correlator is called the tree-level contribution.

Let us now investigate what happens as we let $N\to\infty$.  For example, let us consider the chiral primary operator  $\Psi_L$, rescaled to 
\begin{equation}\label{PsiL}
\Psi_{L}=\frac{(4\pi^2)^{L/2}}{\sqrt{L}N^{L/2}}\Tr Z^L=\frac{(4\pi^2)^{L/2}}{\sqrt{L}N^{L/2}}\,{Z^{A}}_{B}{Z^{B}}_{C}\dots {Z^{\dots}}_{ A}\qquad A,B,C=1..N\,,
\end{equation}
where we have explicitly put in the color indices.  The prefactors are for normalization purposes.  %Since the fields are in the adjoint representation we have that ${Z^A}_A=0$.  
At tree level, the correlator of a
$Z$ field and its conjugate $\overline Z$ is\footnote{We have ignored the fact that  ${Z^A}_A=0$, which is justifiable when we take the large $N$ limit.}
\begin{equation}
\langle {Z^{A}}_{B}(x){{\overline Z}^{\,C }}_D(y)\rangle_{\rm  tree}= \frac{{\delta^A}_D{\delta_{B}}^{C}}{4\pi^2|x-y|^2}\,,
\end{equation}
where we have ignored the fact that  ${Z^A}_A=0$ which is justified in the large $N$ limit.
If we now contract $\Psi_L$ with its conjugate $\overline\Psi_L$, then the leading contribution to the correlator comes from contracting the individual fields in order, as shown in figure 1 (a) and (b).  The contribution of all such ordered contractions is 
\begin{equation}\label{2pt2}
\langle\Psi_{L}(x)\overline \Psi_{L}(y)\rangle_{\rm ordered}= \frac{L N^L}{(\sqrt{L}N^{L/2})^2|x-y|^{2L}}=\frac{1}{|x-y|^{2L}}\,.
\end{equation}
The factor of $N^L$ comes from $L$ factors of ${\delta^{A'}}_A{\delta^A}_{A'}=N$, where each double set of delta functions are from contractions of neighboring fields.  The factor of $L$ comes from the $L$ ways of contracting the fields in the plane, of which (a) and (b) are two examples of this.
\begin{figure}[t]\label{contr}
\centerline{\includegraphics[width=13cm]{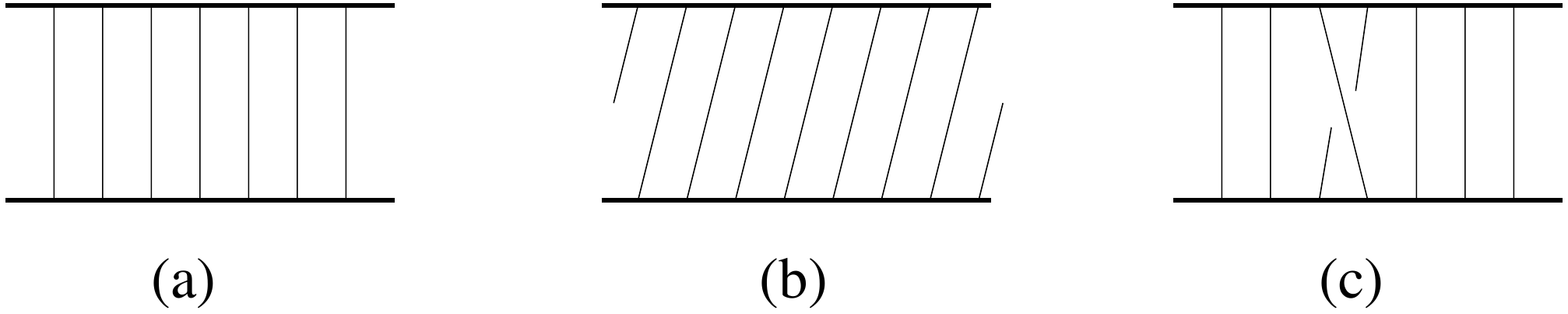}}\caption{Contractions of fields.  The horizontal lines represent the operators and the ordered vertical lines the contractions between the two operators of the individual fields inside the trace. (a) and (b) are planar while (c) is nonplanar.} \nonumber
\end{figure}

Figure 1 (c) is an example of a nonplanar graph, a graph where the lines connecting the fields cannot be drawn in the plane without cutting other lines.  To avoid such cuttings one must lift at least one connecting line out of the plane.  The figure in (c) differs from (a) by two field contractions.  Whereas in (a) we would have had a factor of 
\begin{equation}
\dots{\delta^{A'}}_A{\delta^A}_{A'}{\delta^{B'}}_B{\delta^B}_{B'}{\delta^{C'}}_C{\delta^C}_{C'}\dots=\dots\,N^3\,\dots\,,
\end{equation}
in (c) we have the factor
\begin{equation}
\dots{\delta^{A'}}_A{\delta^A}_{B'}{\delta^{C'}}_B{\delta^B}_{A'}{\delta^{B'}}_C{\delta^C}_{C'}\dots=\dots\,N\,\dots\,,
\end{equation}
where the dots represent contractions that are the same in both cases.  Hence, the nonplanar graph in (c) is suppressed by a factor of $1/N^2$ from that in (a).   In the limit where $N\to\infty$ we can thus ignore this contribution compared to the one in (a) or (b).

All nonplanar graphs will be suppressed by powers of $1/N^2$, where the power depends on the topology of the graph. Actually,  this analysis is valid only if $L<<N$.  If $L$ were on the order of $N$ then the suppression coming from the $1/N$ factors is swamped by the huge number of nonplanar diagrams  compared to the number of planar diagrams.  (There are $L!$ total tree level diagrams of which only $L$ are planar.) 

Generalizing the tree-level correlator in (\ref{2pt2}) to any scalar operator of the form
\begin{equation}
\OO_{I_1,I_2\dots I_L}(x)=\frac{(4\pi^2)^{L/2}}{\sqrt{C_{I_1,I_2\dots I_L}}\,N^{L/2}}\Tr (\phi_{I_1}(x)\phi_{I_2}(x)\dots\phi_{I_L}(x))\,,
\end{equation}
where $C_{I_1,I_2\dots I_L}$ is a symmetry factor (which is $n$ if the indices are invariant when shifting by $L/n$), one finds
\begin{equation}\label{scalartree}
\langle\OO_{I_1,I_2\dots I_L}(x)\overline\OO^{J_1,J_2\dots J_L}(y)\rangle_{\rm tree}=\frac{1}{C_{I_1,I_2\dots I_L}}\left(\delta_{I_1}^{J_1}\delta_{I_2}^{J_2}\dots\delta_{I_L}^{J_L}+\mbox{cycles}\right)\frac{1}{|x-y|^{2L}}\,,
\end{equation}
where ``cycles" refers to the $L-1$ cyclic shifts of the $J_i$ indices.

We next consider the  one-loop contribution to the two-point correlator.  Since we are only considering scalar operators, we only need to consider the the bosonic part of the $\NN=4$ action\footnote{Other parts of the action will contribute at one-loop, but we will show that we can compute their contribution by using an indirect method.}
 which is given by
\begin{equation}\label{action}
S=\frac{1}{2\gym^2}\int d^4x\left\{-\frac12\Tr\FF^2+\Tr\DD_\mu\phi_I\DD^\mu\phi^I-\sum_{I<J}\Tr[\phi_I,\phi_J]^2\right\}\,.
\end{equation}
This action contains a quartic interaction term for the scalars as well as interaction terms between the scalars and the gauge bosons coming from the covariant derivatives.  Hence there will be several types of Feynman graphs that can contribute to the anomalous dimension. 
But because of the robustness of the superconformal algebra, it is sufficient  to only consider  Feynman graphs containing the scalar vertex.  Graphs containing gauge bosons do affect the anomalous dimension, but their contribution can be determined by insisting that chiral primaries have zero anomalous dimension.

If we absorb a factor of $\gym$ into the fields so that their kinetic terms are canonical, then the quartic term can be written as
\begin{equation}\label{4pt}
\frac{\gym^2}{4}\sum_{I,J} \Big(\Tr \phi_I\phi_I\phi_J\phi_J-\Tr\phi_I\phi_J\phi_I\phi_J\Big)\,.
\end{equation}
This vertex should then be inserted in the correlator and be Wick contracted with two neighboring fields in the incoming operator and two neighboring fields in the outgoing operator so that the resulting Feynman graph is planar.  This is shown in figure 2.  
\begin{figure}[t]\
\centerline{\includegraphics[width=12cm]{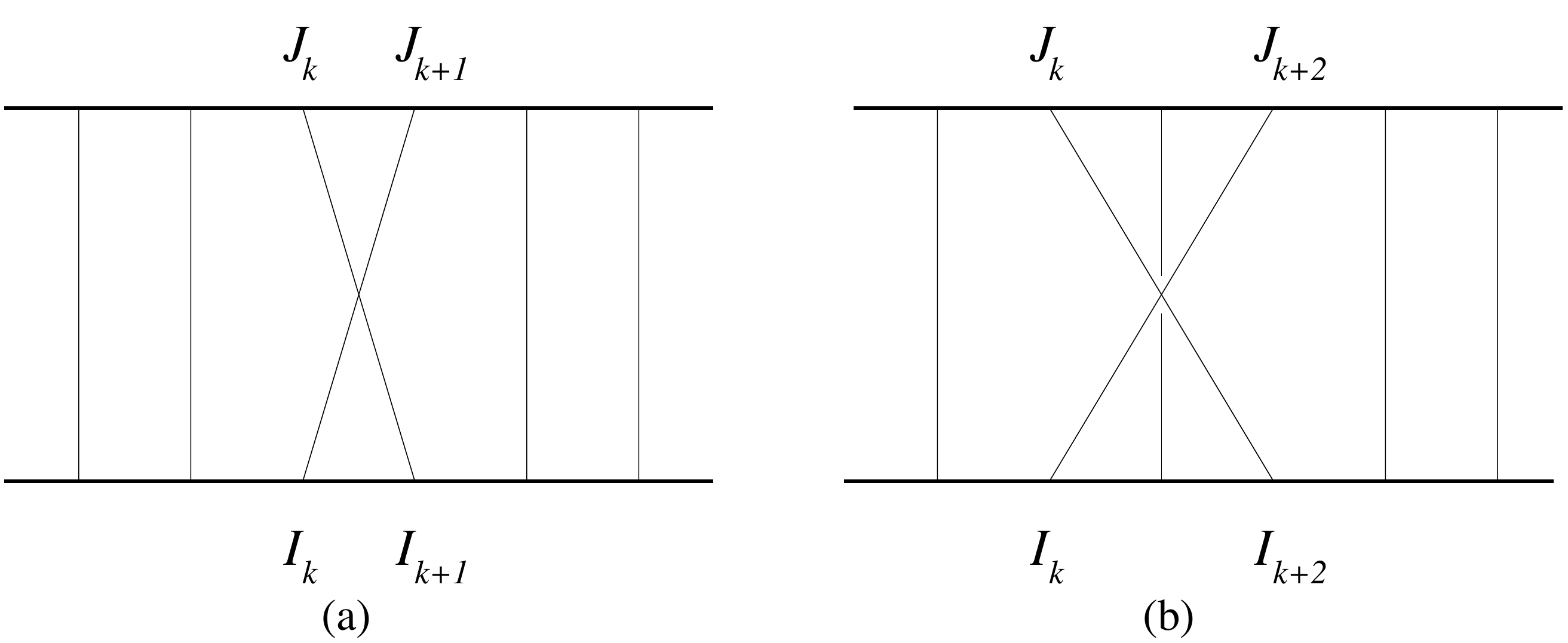}}\caption{Quartic interaction inserted into the correlator, connecting (a) two neighboring fields  (b) nonneighboring fields.  Case (b) is nonplanar.  Notice that the interaction has added a loop to the diagrams. } \nonumber
\end{figure}
In particular, we should consider the  subcorrelator from (\ref{scalartree}),
\begin{eqnarray}\label{corr1l}
&&\Big\langle{(\phi_{I_k}\phi_{I_{k+1}})^A}_{C}(x)\left(\frac{i\,\gym^2}{4}\int d^4z\sum_{I,J} (\Tr \phi_I\phi_I\phi_J\phi_J(z)-\Tr\phi_I\phi_J\phi_I\phi_J(z))\right)\nonumber\\
&&\qquad\qquad\qquad\qquad\qquad\qquad\times{(\phi^{J_{k+1}}\phi^{J_k})^{C'}}_{A'}(y)\Big\rangle\nonumber\\
&&=i\,\frac{N}{(4\pi^2)^2}{\delta^A}_{A'}{\delta_{C}}^{C'}\frac{\gym^2N}{64\,\pi^4}\left(2{\delta_{I_k}}^{J_k}{\delta_{I_{k+1}}}^{J_{k+1}}+2\delta_{I_kI_{k+1}}\delta^{J_kJ_{k+1}}-4{\delta_{I_k}}^{J_{k+1}}{\delta_{I_{k+1}}}^{J_k}\right)\nonumber\\
&&\qquad\qquad\qquad\qquad\qquad\qquad\times \int \,\frac{d^4z}{|z-x|^4|z-y|^4}\,.
\end{eqnarray}
%where we have picked up an extra color factor that we have grouped with the $\gym^2$.
The set of delta functions for the flavor indices arise from the two terms in (\ref{4pt}).  There are four planar ways to contract the indices in (\ref{4pt}) with the incoming and outgoing fields.    The first term either contracts the incoming indices with  the outgoing indices in order, or it contracts incoming to incoming and outgoing to outgoing.  The second term in (\ref{4pt}) always contracts the indices between the incoming and outgoing fields in reverse order.
Note that there are two factors of $N$ in (\ref{corr1l}), coming from sums over color factors, while the correlator
$\langle{(\phi_{I_k}\phi_{I_{k+1}})^A}_{C}(x){(\phi^{J_{k+1}}\phi^{J_k})^{C'}}_{A'}(y)\rangle$ has only one such factor.  In fact, it is not difficult to see that for all planar graphs, every factor of $\gym^2$ comes with a factor of $N$.  Hence, it is convenient to define  the 't Hooft coupling, $\lambda\equiv\gym^2N$, as a new expansion parameter.   

The integral in (\ref{corr1l}) has a logarithmic divergence as $z\to x$ and $z\to y$, hence it is necessary to add a UV cutoff $\Lambda$.   There is no IR divergence since the integral is well behaved as $z\to\infty$.  The integral over $z$ is in Minkowski space, but it can be Wick rotated to Euclidean space such that $d^4z\to id^4z_E$.  With the UV cutoff the integral is restricted to the region where  $|z_E-x|\ge\Lambda^{-1}$ and $|z_E-y|\ge\Lambda^{-1}$.   The integral is then dominated by the regions near the cutoff and can be approximated to
\begin{equation}
i\, \int \,\frac{d^4z_E}{|z-x|^4|z-y|^4}\approx \frac{2\,i}{|x-y|^4}\,\int_{\Lambda^{-1}}^{|x-y|}\frac{d\xi\,d\Omega_3}{\xi}=\frac{2\,\pi^2\,i}{|x-y|^4}\,\ln(\Lambda^2|x-y|^2)\,.
\end{equation}
Therefore the subcorrelator in (\ref{corr1l}) becomes
\begin{eqnarray}\label{1loopx}
\frac{N{\delta^A}_{A'}{\delta_{C}}^{C'}}{(4\pi^2)^2|x-y|^4}\frac{\lambda}{16\,\pi^2}\left(2\delta_{I_kI_{k+1}}\delta^{J_kJ_{k+1}}-{\delta_{I_k}}^{J_{k+1}}{\delta_{I_{k+1}}}^{J_k}-{\delta_{I_k}}^{J_k}{\delta_{I_{k+1}}}^{J_{k+1}}\right)\,\ln(\Lambda^2|x-y|^2)\,.\nonumber\\
\end{eqnarray}

Normally one does loop integrals in momentum space.  We could have done that here as well, but for these particular one-loop calculations it is easier to do things in coordinate space.  This is mainly because the operators are local so all fields within the operator are at the same coordinate position, simplifying the calculation.

There will also be other one-loop contributions to the correlators.  Figure 3 shows some examples of these. These can come from gluon exchange between scalar fields or self energy diagrams.  We could compute these contributions explicitly, but we will soon show that we do not actually need to do this.  At this point we note that since the $R$-charge is conserved and since gluons have no $R$-charge% and since the flavor of a scalar going into a self-energy diagram is the same coming out
, then these types of diagrams %will not change the flavor indices.  Hence, these interactions  
will only lead to terms where all incoming indices  are contracted sequentially with the outgoing indices, giving the same flavor structure as the planar tree-level graphs.  
\begin{figure}[t]\
\centerline{\includegraphics[width=13cm]{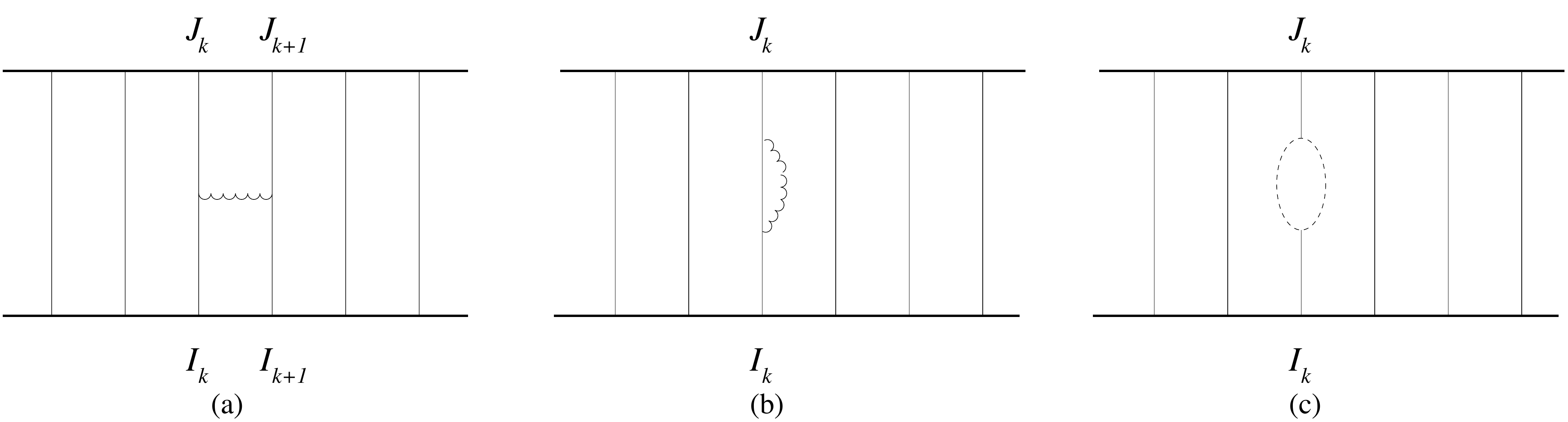}}\caption{One-loop planar graphs that do not affect the flavor structures. (a) A gluon exchange between neighboring scalars.  The gluon carries no $R$-charge, so the flavor indices are unchanged. (b) Scalar self-energy from a gluon.    (c) Scalar self-energy from a fermion loop.   $R$-charge conservation  and the fact that only one scalar line is involved means that (b) and (c) leave the flavor indices unchanged.} \nonumber
\end{figure}

Applying these arguments to  the correlator in (\ref{scalartree}), we find the one-loop result
%. $\langle\OO_1(x)\OO_2(y)\rangle$, where the $\OO(x)$ are in the $SO(6)$ sector, that is they are composed of $L$ scalar fields with explicit flavor indices:
%\begin{eqnarray}
%\OO_1(x)=\frac{1}{N^{L/2}}\Tr(\phi_{i_1}(x)\dots\phi_{i_L}(x))\,\qquad
%\OO_2(x)=\frac{1}{N^{L/2}}\Tr(\phi_{j_1}(x)\dots\phi_{j_L}(x))\,,
%\end{eqnarray}
%Examining (\ref{1loopx}), we see that the one loop contribution to the correlator is 
\begin{eqnarray}\label{one-loop}
&&\langle\OO_{I_1,I_2\dots I_L}(x)\overline\OO^{J_1,J_2\dots J_L}(y)\rangle_{\mbox{one-loop}}\nonumber\\
&&=\frac{\lambda}{16\pi^2}\frac{\ln(\Lambda^2|x-y|^2)}{|x-y|^{2L}}
\sum_{\ell=1}^L(2P_{\ell,\ell+1}-K_{\ell,\ell+1}-1+C)\frac{1}{\sqrt{C_{I_1,\dots I_L}C_{J_1,\dots J_L}}}\,\,\delta_{I_1}^{J_1}\delta_{I_2}^{J_2}\dots\delta_{I_L}^{J_L}\nonumber\\
&&\qquad\qquad \qquad\qquad +\ {\rm cycles}\,.
\end{eqnarray}
There is a  sum over $\ell$ because the diagram in figure 2(a) can have the interaction between any of the $L$ pairs of neighboring fields.
The constant $C$ comes from the diagrams in figure 3.  ``Cycles" again refers to the $L-1$ uniform shifts of the $J_k$ indices.

$P_{\ell,\ell+1}$ is the exchange operator, and as its name implies it exchanges the flavor indices of the $\ell$ and the $\ell+1$ sites inside the trace.  Its action on the  $\delta$-functions in (\ref{one-loop}) is
\begin{equation}
P_{\ell,\ell+1}\,\delta_{I_1}^{J_1}\dots\delta_{I_\ell}^{J_\ell}\delta_{I_{\ell+1}}^{J_{\ell+1}}\dots\delta_{I_L}^{J_L}=\delta_{I_1}^{J_1}\dots\delta_{I_\ell}^{J_{\ell+1}}\delta_{I_{\ell+1}}^{J_{\ell}}\dots\delta_{I_L}^{J_L}\,.
\end{equation}
$K_{\ell,\ell+1}$ is the  trace operator which contracts the flavor indices of neighboring fields.  Its  action on the $\delta$-functions is
\begin{equation}
K_{\ell,\ell+1}\,\delta_{I_1}^{J_1}\dots\delta_{I_\ell}^{J_\ell}\delta_{I_{\ell+1}}^{J_{\ell+1}}\dots\delta_{I_L}^{J_L}=\delta_{I_1}^{J_1}\dots\delta_{I_\ell I_{\ell+1}}\delta^{{J_\ell}J_{\ell+1}}\dots\delta_{I_L}^{J_L}\,.
\end{equation}
Because of the $P_{\ell,\ell+1}$ and $K_{\ell,\ell+1}$ there is operator mixing at the one-loop level.

Adding the one-loop correlator  to the tree level correlator in (\ref{scalartree}) we get the expression
\begin{eqnarray}
&&\langle\OO_{I_1,I_2\dots I_L}(x)\overline\OO^{J_1,J_2\dots J_L}(y)\rangle=\nonumber\\
&&\quad\frac{1}{|x-y|^{2L}}\left(1-\frac{\lambda}{16\pi^2}{\ln(\Lambda^2|x-y|^2)}
\sum_{\ell=1}^L(C-1-2P_{\ell,\ell+1}+K_{\ell,\ell+1})\right){\delta^{j_1}}_{i_1}\dots
{\delta^{j_L}}_{i_L}\nonumber\\
&&\qquad\qquad +\ {\rm cycles}\,.
\end{eqnarray}
If we compare this result to (\ref{2pta}), we see that because of  the operator mixing  the anomalous dimension $\gamma$ should be replaced with an  operator, $\Gamma$, where
%whose eigenvalues are $\gamma$.  In particular, we find for operators made up of scalar fields, that $\Gamma$ is given by \cite{Minahan:2002ve}
\begin{equation}\label{Gammaeq}
\Gamma=\frac{\lambda}{16\pi^2}\sum_{\ell=1}^L(1-C-2P_{\ell,\ell+1}+K_{\ell,\ell+1})\,.
\end{equation}
The possible one-loop anomalous dimensions are then found by diagonalizing $\Gamma$. 

The entire class of scalar single trace operators of length $L$ can be mapped to a Hilbert space which itself is a tensor product of finite dimensional Hilbert spaces
\begin{equation}
\VV_1\otimes\VV_2\dots\otimes\VV_\ell\otimes\dots\otimes\VV_L\,.
\end{equation}
Each $\VV_\ell$ is the Hilbert space for an $SO(6)$ vector representation, {\it i.e.} $CP^5$.   The tensor product is the same  Hilbert space as that of a one-dimensional spin-chain with $L$ sites, where at each site  there is an $SO(6)$ vector ``spin" (see figure 4).  Because of the cyclicity property of the trace, we should include the further restriction that the Hilbert space be invariant under the shift
\begin{equation}\label{shift}
\VV_1\otimes\VV_2\dots\otimes\VV_\ell\otimes\dots\otimes\VV_L\to\VV_L\otimes\VV_1\dots\otimes\VV_{\ell-1}\otimes\dots\otimes\VV_{L-1}\,.
\end{equation}

The operator $\Gamma$ in (\ref{Gammaeq}) acts linearly on this space:
\begin{equation}
\Gamma:\VV_1\otimes\VV_2\dots\otimes\VV_\ell\otimes\dots\otimes\VV_L\to \VV_1\otimes\VV_2\dots\otimes\VV_\ell\otimes\dots\otimes\VV_L\,.
\end{equation}
Furthermore, it is Hermitian and commutes with the shift in (\ref{shift}).  Thus, we can treat $\Gamma$ as a Hamiltonian on the spin-chain.    The energy eigenstates then correspond to the possible anomalous dimensions for the scalar operators. Since the Hamiltonian commutes with the shift, it is also consistent to project onto eigenstates that are invariant under the shift.   Because $P_{\ell,\ell+1}$ and $K_{\ell,\ell+1}$ act on neighboring fields, the spin-chain Hamiltonian only has nearest neighbor interactions between the spins.
\begin{figure}[t]
\centerline{\includegraphics[width=13cm]{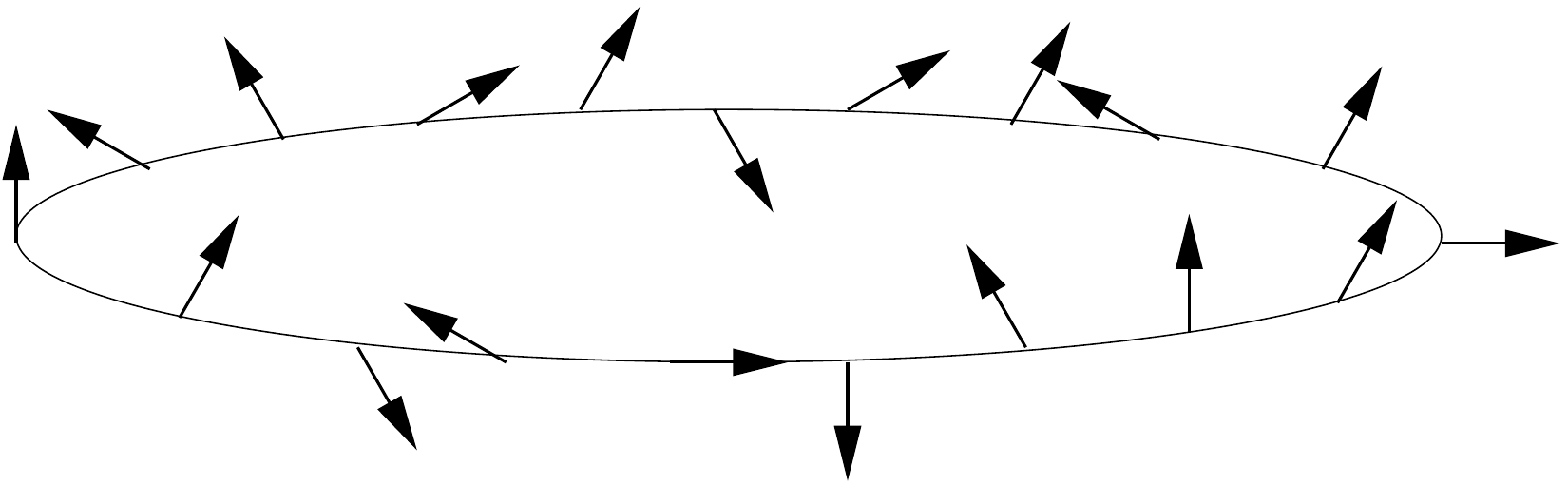}}\caption{A spin-chain with $SO(6)$ vector sites.} \nonumber
\end{figure}

One particular eigenstate of $\Gamma$ corresponds to the chiral primary $\Psi_L$ in (\ref{PsiL}).  $\Psi_L$ is symmetric under the exchange of any field, hence $P_{\ell,\ell+1}\Psi_L=\Psi_L$ for any $\ell$.  Furthermore, $\Psi_L$ has only $Z$ fields and not $\overline Z$ fields, thus $K_{\ell,\ell+1}\Psi_L=0$.  This generalizes to any chiral primary, which is in the $L^{\rm th}$ symmetric traceless representation of $SO(6)$.  Therefore,
\begin{equation}
\Gamma\,\Psi_L=\frac{\lambda}{16\,\pi^2}\sum_{\ell=1}^L(1-C-2)\Psi_L
\end{equation}
However, the dimension of $\Psi_L$ is protected, meaning that its anomalous dimension is zero.  Hence, we find that $C=-1$ and $\Gamma$ becomes \cite{Minahan:2002ve}
\begin{equation}\label{Gammaeqfinal}
\Gamma=\frac{\lambda}{8\,\pi^2}\sum_{\ell=1}^L\left(1-P_{\ell,\ell+1}+\frac{1}{2}\,K_{\ell,\ell+1}\right)\,.
\end{equation}

Another useful way to write $\Gamma$ is in terms of projectors.  The tensor product of two $SO(6)$ vector representations is reducible into the traceless symmetric, the antisymmetric, and the singlet representations.  The operators that project $\VV_\ell\otimes\VV_{\ell+1}$ onto these three representations are
\begin{equation}
\Pi_{\ell,\ell+1}^{\rm sym}=\frac{1}{2}(1+P_{\ell,\ell+1})-\frac{1}{6}K_{\ell,\ell+1}\,,\ \ 
\Pi_{\ell,\ell+1}^{\rm as}=\frac{1}{2}(1-P_{\ell,\ell+1})\,,\ \ 
\Pi_{\ell,\ell+1}^{\rm sing}=\frac{1}{6}K_{\ell,\ell+1}\,.
\end{equation}
We can then write $\Gamma$ as
\begin{equation}\label{Gammaeqproj}
\Gamma=\frac{\lambda}{8\,\pi^2}\sum_{\ell=1}^L\left(0\,\Pi_{\ell,\ell+1}^{\rm sym}+2\,\Pi_{\ell,\ell+1}^{\rm as}+3\,\Pi_{\ell,\ell+1}^{\rm sing}\right)\,,
\end{equation}
with only two of the three projectors contributing to $\Gamma$.

Although we will not show it here, the Hamiltonian that corresponds to $\Gamma$ for the spin-chain is integrable \cite{Minahan:2002ve}.  There is a precise meaning for what this means which will be explained in later chapters of the review (see \cite{chapABA,chapSMat,chapYang}).  For us it means that the system is solvable, at least  in principle.  We will give a taste of this in the next section where we consider a certain subset of scalar operators.  

Going beyond  one-loop, one finds that the $n$-loop contribution to the anomalous dimension can involve up to  $n$ neighboring fields in an effective Hamiltonian \cite{Beisert:2003tq,Beisert:2004ry} (see  \cite{chapHigher}).  Therefore, as $\lambda$ becomes larger these longer range interactions become more and more important, such that at strong coupling the spin-chain is effectively long range.  In this case the Hamiltonian is not known above the first few loop orders \cite{Beisert:2003tq,Beisert:2004ry,Sieg:2010tz}.

\section{One-loop generalization to all single trace operators}

In this subsection we describe the generalization of $\Gamma$ to all single trace operators.  We do not give a derivation here, but instead refer the reader to the references.

In the general case the ``spins" at each site of the chain are made up of the elements of the singleton representation enumerated in (\ref{singleton}).  The Hilbert space is then the tensor product in (\ref{tensorprod}) projected onto states invariant under the shift in (\ref{shift1}).  The one-loop anomalous dimension is then described by a Hamiltonian with nearest neighbor interactions.
Unlike the scalar case where the spins are in a finite dimensional representation, the singleton representation is infinite dimensional.  However, there is still a beautiful way to write the Hamiltonian in terms of projectors \cite{Beisert:2003jj,Beisert:2003yb}.  

The various $PSU(2,2|4)$ representations can be expressed in terms of their highest weights which are given by the six charges of the $PSU(2,2|4)$ Cartan subalgebra.  The singleton is then labeled by $(1,0,0;1,0,0)$, where the highest weight in the representation belongs to the $Z$ field.  The Hamiltonian will involve the tensor product of two singleton representations which decomposes as
\begin{equation}
\VV\otimes\VV=\sum_{j=0}^\infty\VV_j\,.
\end{equation}
The first two representations in this decomposition are different from the others and have the highest weights
\begin{equation}
\VV_0:\ \ (2,0,0;2,0,0)\,\qquad \VV_1:\ \ (2,0,0;1,1,0)\,.
\end{equation}
The other representations have highest weights
\begin{equation}
\VV_j:\ \ (j,j-2,j-2;0,0,0)\qquad j\ge 2\,.
\end{equation}
Notice that if we limit ourselves to scalar fields with no Lorentz charges then the only representations in play are $\VV_0$, $\VV_1$ and $\VV_2$, whose decompositions under the $SO(6)$ subgroup contain the   symmetric traceless, the antisymmetric, and singlet representations respectively.

The Hamiltonian for the complete spin-chain has the compact form \cite{Beisert:2003jj,Beisert:2003yb}
\begin{equation}\label{Gammafull}
\Gamma=\frac{\lambda}{8\,\pi^2}\sum_{\ell=1}^L\sum_{j=0}^\infty\,2\,h(j)\,\Pi_{\ell,\ell+1}^{(j)}\,,
\end{equation}
where $\Pi_{\ell,\ell+1}^{(j)}$ projects $\VV_{\ell}\otimes\VV_{\ell+1}$ onto $\VV_j$ and $h(j)$ is the harmonic sum defined by\footnote{In \cite{Lipatov:1997vu} Lipatov remarked that harmonic sums would appear in the anomalous dimension matrix for $\NN=4$ SYM, leading him to predict that the theory would be solvable.}
\begin{equation}
h(j)\equiv\sum_{k=1}^j\frac{1}{k}\,.
\end{equation}
Examining the expression for  $\Gamma$ in (\ref{Gammaeqproj}) we see that it has the form in (\ref{Gammafull}) when only $j=0,1,2$ contribute.

\section{Closed sectors }
Since we have operator mixing, the alert reader could very well be concerned that scalar field operators will  mix with operators that contain non-scalar fields.  In turns out that generally this can happen, but not at the one-loop level.  

Operator mixing preserves the total charges of the $PSU(2,2|4)$ symmetry group.  This is because the  anomalous dimension matrix  is the  the dilatation operator $D$ minus the bare dimension.  To see why this matters consider the complete dilatation operator, which can be expressed as an expansion in $\lambda$ of the form
\begin{equation}
D=\sum_{n=0}^\infty \lambda^n\,D^{(2n)}\,.
\end{equation}
$D^{(0)}$ gives the bare dimension of the operator while $D^{(2)}$ is the one-loop anomalous dimension operator $\Gamma$ in (\ref{Gammaeqfinal}) for scalar single trace operators or (\ref{Gammafull}) for the most general single trace operators.  The dilation operator commutes with the Lorentz generators and the $R$-symmetry generators.  Since this is true for any value of $\lambda$, it must be true that all $D^{(2n)}$ commute with these generators.  Hence, the Lorentz and $R$-charges are preserved by the mixing.  Furthermore, each of the $D^{(2n)}$ commutes with $D^{(0)}$, which can be established by power counting in the graphs.  Therefore, mixing only occurs between operators with the same $R$-charges, Lorentz charges, and bare dimensions.

We can use this information to show the existence of closed sectors.  One such sector  are operators made up of two types of scalar fields, say, $Z$ and $W$, which have the charges $(1,0,0;1,0,0)$ and $(1,0,0;0,1,0)$ respectively.  Hence, the total charges of a single trace operator made up of $L-M$ $Z$ fields and $M$ $W$ fields is $(L,0,0;L-M,M,0)$.  The mixing must preserve these charges and the only way to do this is to mix with operators having the same number of $Z$ and $W$ fields with possible rearrangements to their order, as one can verify by checking the charges for the other fields.  This closed sector is called the $SU(2)$ sector, since $Z$ and $W$ make up a doublet of an $SU(2)$ subgroup of the  $R$-symmetry group.

If we  now include a third type of scalar field $X$, then the combination $ZW\!X$ which has charges $(3,0,0;1,1,1)$ can mix with two fermions with individual charges $(\frac{3}{2},\frac12,0;\frac12,\frac12,\frac12)$ and $(\frac{3}{2},-\frac12,0;\frac12,\frac12,\frac12)$ but is otherwise closed \cite{Beisert:2003ys}.  This closed sector is the $SU(2|3)$ sector containing an $SU(3)$ subgroup of the $R$-symmetry and an $SU(2)$ subgroup of the Lorentz group.  The scalars make up a triplet of the $SU(3)$ and are singlets under the $SU(2)$ while the fermions are singlets under the $SU(3)$ and make up a doublet of the $SU(2)$.  Notice that this sort of mixing changes the number of fields in the trace.  Such mixing is called dynamical \cite{Beisert:2003ys}.

We call the full set of scalar operators the $SO(6)$ sector since the fields form a representation of the full $R$-symmetry group but are singlets under the Lorentz group.  However, this cannot be a closed sector, since not even an $SU(3)$ subsector is closed to mixing with fields with non-zero Lorentz charges.  In fact, the $SO(6)$ sector can mix into operators containing any one of the fields so the smallest closed sector containing $SO(6)$ is the full $PSU(2,2|4)$.  However, the mixing outside of the $SO(6)$ sector is dynamical, but dynamical mixing cannot occur until the two-loop level  \cite{Beisert:2003ys}.  Hence the $SO(6)$ sector is closed  at one-loop.

Both $SU(2)$ and $SU(2|3)$ are compact groups and so  the fields in these closed sectors are part of a finite dimensional representation of the group.  There is another important closed sector where this is not the case.  This is the $SU(1,1)$ sector (also called the $SL(2)$ sector) \cite{Beisert:2003yb}.  In this sector we only have one type of scalar field, say $Z$, and  covariant derivatives with one type of polarization, say $\DD_{++}$ which has charges $(1,\frac12,\frac12;0,0,0)$.  A typical single trace operator in this sector could have $L$ scalar fields and $M$ covariant derivatives.  The mixing occurs by redistributing the $M$ covariant derivatives among the $L$ fields.  Notice that this sector is nondynamical.  Notice further that the fields fall into an infinite dimensional representation of $SU(1,1)$ since we can have an arbitrary number of covariant derivatives on any $Z$ field.  In fact the $SU(1,1)$ sector even appears in QCD \cite{Lipatov:1993yb,Faddeev:1994zg} (see \cite{chapQCD}).

\section{The \texorpdfstring{$SU(2)$}{SU(2)} sector and the Heisenberg spin-chain}

Let us now restrict our single trace operators to the $SU(2)$ closed sector.  The two independent fields transform under a doublet of $SU(2)$, hence we can label the $Z$ field as spin up ($\up$)  and the $W$ field as spin down ($\down$). There is no contribution from  $K_{\ell,\ell+1}$ in (\ref{Gammaeqfinal}) since the operators only have $Z$ and $W$ fields and not their conjugates.  Thus, the $SU(2)$ sector has the Hamiltonian
\begin{equation}\label{SU2ham}
\Gamma_{SU(2)}=\frac{\lambda}{8\pi^2}\sum_{\ell=1}^L(1-P_{\ell,\ell+1})\,.
\end{equation}

In terms of spin operators the Hamiltonian can be rewritten as
\begin{equation}\label{spchham}
\Gamma_{SU(2)}=\frac{\lambda}{8\pi^2}\sum_{\ell=1}^L\left(\frac12-2\,\vec S_\ell\cdot\vec S_{\ell+1}\right)\,.
\end{equation}
Remarkably, $\Gamma_{SU(2)}$ is the Hamiltonian of the Heisenberg spin-chain with $L$ lattice sites.  The total spin $\vec S=\sum_\ell \vec S_\ell$ commutes with $\Gamma$ so the energy eigenstates are simultaneously total spin eigenstates.  This should not be surprising since we have already established that the dilatation operator commutes with the $R$-symmetry  and the spin here is one of its subgroups.

Because of the sign of the $\vec S_\ell\cdot\vec S_{\ell+1}$ term the spin-chain is ferromagnetic and the ground state has all spins  aligned, with total spin $L/2$.  This is the symmetric representation, which  corresponds to the chiral primary operator.  A quick check of the Hamiltonian in (\ref{spchham}) shows that its energy is zero.  The operators which are not chiral primaries correspond to  excitations about the ground state.  They have total spin that is less than $L/2$.
A full description on how to find these other states is given in \cite{Minahan:2002ve}.  Here we give a partial description based on an $S$-matrix approach (see \cite{chapABA,chapSMat}). 

 Let us start with a ground state which we write as $|\up\up\up\dots\up\up\rangle$.  This corresponds to the chiral primary $\Psi_L$ described in an earlier section.  Let us now consider the states where one spin is down.  In this case the Hamiltonian in (\ref{SU2ham}) acts like a constant plus a  hopping term, moving the down spin either one site to the left or the right.  In particular, the action on a state with a down spin at a particular position $\ell$ is 
 \begin{eqnarray}
 &&\Gamma_{SU(2)}|\up\dots\up\stackrel{\ell}\down\up\dots\up\rangle\nonumber\\
 &&\qquad=\frac{\lambda}{8\,\pi^2}\left(2\,|\up\dots\up\stackrel{\ell}\down\up\dots\up\rangle-|\up\dots\stackrel{\ell-1}\down \up \up\dots\up\rangle-|\up\dots\up\up\stackrel{\ell+1}\down\dots\up\rangle\right)\,.\nonumber\\
 \end{eqnarray}
 From this it is easy to see that the  eigenstates are
\begin{equation}
|p\rangle\equiv\frac{1}{\sqrt{L}}\sum_{\ell=1}^L e^{ip\ell}|\up\up\dots\stackrel{\ell}{\down}\dots\up\up\rangle
\end{equation}
where
\begin{equation}
\Gamma_{SU2}|p\rangle=\ve(p)\,|p\rangle\,,\qquad \ve(p)=\frac{\lambda}{2\,\pi^2}\,\sin^2\frac{p}{2}\,.
\end{equation}
The state $|p\rangle$ is called a single magnon state with momentum $p$.  The dispersion is $\ve(p)$ and the magnon momentum $p$ must be quantized so that the state is invariant under the shift $\ell\to \ell+L$, therefore $p=2\pi n/L$.  If $n=0$ then this is the symmetric state and so this has total spin $L/2$.  All other cases have  total spin $L/2-1$.  This is fine for an ordinary spin chain, but we must remember that our states need to be invariant under the shift $\ell\to\ell+1$ since the single trace operators are invariant if we shift all fields over by one position.  Hence, the only allowed state is the $p=0$ state and we find no operators that are not chiral primaries with only a single $W$ field.

The first nontrivial case occurs with two down spins, since here it will be possible to satisfy the trace condition but not be in the symmetric representation.  We will construct these states using an argument that goes back to Yang and Yang \cite{Yang:1966ty}.  Instead of a closed chain of length $L$ let us suppose we have a chain of infinite length.  Consider the unnormalized two magnon state
\begin{equation}\label{2magnon}
|p_1,p_2\rangle=\sum_{\ell_1<\ell_2}e^{ip_1\ell_1+ip_2\ell_2}|\dots\stackrel{\ell_1}\down\dots\stackrel{\ell_2}\down\dots\rangle+e^{i\phi}\sum_{\ell_1>\ell_2}e^{ip_1\ell_1+ip_2\ell_2}|\dots\stackrel{\ell_2}\down\dots\stackrel{\ell_1}\down\dots\rangle\,,
\end{equation}
where we assume that $p_1>p_2$.  We can think of $|p_1,p_2\rangle$ as the scattering state for two magnons.  The first term is the incoming part  while the second term is the outgoing part.  The phase $e^{i\phi}$ is then the $S$-matrix $S_{12}$ for the scattering. It is clear that if $|p_1,p_2\rangle$ is to be an eigenstate of $\Gamma_{SU(2)}$ then the eigenvalue will be the sum of the eigenvalues of two single magnon states with magnon momenta $p_1$ and $p_2$ respectively, since for $|\ell_1-\ell_2|>>1$ the two magnons cannot be interacting with each other.  The subtlety occurs when the two down spins are next to each other, because the Hamiltonian cannot hop a down spin on top of another down spin.  However, by adjusting the phase $e^{i\phi}$ we can ensure that $|p_1,p_2\rangle$ is an eigenstate.  If we concentrate on all the ways the Hamiltonian puts the  two down spins next to each other at sites $\ell$ and $\ell+1$ we find that in order to have an eigenstate we must satisfy the  equation
\begin{eqnarray}
&&e^{ip_2}\left(2-e^{-ip_1}-e^{ip_2}\right)+e^{ip_1}\left(2-e^{ip_1}-e^{-ip_2}\right)e^{i\phi}\nonumber\\
&&\qquad=\left(4-e^{-ip_1}-e^{ip_1}-e^{-ip_2}-e^{ip_2}\right)\left(e^{ip_2}+e^{ip_1}e^{i\phi}\right)\,,
\end{eqnarray}
which has the solution
\begin{equation}\label{Smatrix}
e^{i\phi}=S_{12}=-\,\frac{e^{ip_1+ip_2}-2e^{ip_2}+1}{e^{ip_1+ip_2}-2e^{ip_1}+1}
\end{equation}

Now let us put the two magnons back on a cyclic spin chain of length $L$.  The trace condition enforces the total momentum to be $p_1+p_2=0$.  The quantization condition for $p_1$ works as follows.  If we transport the magnon once around the circle the state is invariant. However, the transport brings the first magnon past the second one, so it also picks up a phase $e^{i\phi}$.  Hence we have that $e^{ip_1L}e^{i\phi}=1$.  With $p_2=-p_1$ we readily see that $e^{i\phi}=e^{-ip_1}$.  Thus the allowed values for $p_1$ are $p_1=2\pi n/(L-1)$ and the possible eigenvalues for the two magnon state are 
$$\gamma=\frac{\lambda}{\pi^2}\sin^2\frac{\pi n}{L-1}\,.$$
The case where $n=0$ is the symmetric state with spin $L/2$.  All other choices have spin $L/2-2$.

To go even further, it is convenient to define the rapidity variable $u$, where $e^{ip}=\frac{u+i/2}{u-i/2}$.  The dispersion relation is then
\begin{equation}\label{dispersion}
\ve(u)=\frac{\lambda}{8\,\pi^2}\,\frac{1}{u^2+1/4}\,,
\end{equation}
while the $S$-matrix in (\ref{Smatrix}) for magnons with rapidity variables $u_j$ and $u_k$ is
\begin{equation}
S_{jk}=\frac{u_j-u_k-i}{u_j-u_k+i}\,.
\end{equation}
For $M$ magnons one then sets up a state
\begin{equation}
|p_1,p_2,\dots p_M\rangle=\sum_{\ell_1<\ell_2\dots\ell_M}e^{ip_1\ell_1+ip_2\ell_2+\dots+ip_M\ell_M}|\dots\stackrel{\ell_1}{\down}\dots\stackrel{\ell_2}{\down}\dots\dots\stackrel{\ell_M}{\down}\dots\rangle + \dots\,
\end{equation}
with $p_1>p_2\dots>p_M$ and where the last set of dots refers to the other possible orderings for the magnons, with appropriate phase factors.  One can show that the phase factors are products of the two-particle $S$-matrices, which makes the system integrable.  Putting the magnons on a circle with $L$ sites we then find the quantization condition for the $j^{\rm th}$ magnon
 \begin{equation}\label{bethe}
 \left(\frac{u_j+i/2}{u_j-i/2}\right)^L=\prod_{k\ne j}^M\frac{u_j-u_k+i}{u_j-u_k-i}\,.
 \end{equation}
 The energy of the state is
 \begin{equation}
 \gamma=\sum_{j=1}^M\ve(u_j),
 \end{equation}
 where $\ve(u_j)$ is given by (\ref{dispersion}).
 The trace condition for the total momentum is
 \begin{equation}
 \prod_{j=1}^M \frac{u_j+i/2}{u_j-i/2}=1
 \end{equation}
 
 The equations in (\ref{bethe}) were first derived by Bethe many years ago\cite{Bethe:1931hc} and are called the Bethe equations for the Heisenberg spin chain.  Further solutions to these equations can be found in \cite{Minahan:2002ve,Beisert:2003xu,Beisert:2003ea}.   Their  generalization to other sectors including the full $PSU(2,2|4)$ long-range spin chain \cite{Beisert:2005fw} are discussed in 
 \cite{chapLR,chapCurve,chapSProp}

\bigskip
\noindent {\bf Acknowledgments}:   This work was supported in part by Vetenskapr\aa det.

%%%%%%%%%%%%%%%%%%%%%%%%%%%%%%%%%%%%%%%%
%\phantomsection
\addcontentsline{toc}{section}{\refname}
%\bibliography{references}
\bibliography{references,chapters}
\bibliographystyle{nb}

\end{document}